\documentclass[12pt,preprint]{aastex}

\usepackage{graphicx}

\newcommand{\mycite}[1]{[\cite{#1}]}

\shorttitle{Architecture and Algorithms for Kepler Error Correction}
\shortauthors{M. C. Stumpe et al.}

\begin{document}


\title{\textit{Kepler} Presearch Data Conditioning {I} -- Architecture and Algorithms for Error Correction in \textit{Kepler} Light Curves}

\author{Martin C. Stumpe\altaffilmark{*,1,2}, Jeffrey C. Smith\altaffilmark{1,2}, Jeffrey E. Van Cleve\altaffilmark{1,2}, Joseph D. Twicken\altaffilmark{1,2}, Thomas S. Barclay\altaffilmark{1,3}, Michael N. Fanelli\altaffilmark{1,3}, Forrest R. Girouard\altaffilmark{1,4}, Jon M. Jenkins\altaffilmark{1,2}, Jeffery J. Kolodziejczak\altaffilmark{5}, Sean D. McCauliff\altaffilmark{1,4}, Robert L. Morris\altaffilmark{1,2}}

\altaffiltext{*}{email: martin.stumpe@nasa.gov}
\altaffiltext{1}{NASA Ames Research Center, Moffett Field, CA, 94035}
\altaffiltext{2}{SETI Institute}
\altaffiltext{3}{Bay Area Environmental Research Institute}
\altaffiltext{4}{Orbital Sciences Corporation}
\altaffiltext{5}{Marshall Space Flight Center, Huntsville, AL 35812}


\newpage \clearpage

\begin{abstract}
\textit{Kepler} provides light curves of 156,000 stars with unprecedented precision. However, the raw data as they come from the spacecraft contain significant systematic and stochastic errors. These errors, which include discontinuities, systematic trends, and outliers, obscure the astrophysical signals in the light curves. To correct these errors is the task of the Presearch Data Conditioning (PDC) module of the \textit{Kepler} data analysis pipeline. The original version of PDC in \textit{Kepler} did not meet the extremely high performance requirements for the detection of miniscule planet transits or highly accurate analysis of stellar activity and rotation. One particular deficiency was that astrophysical features were often removed as a side-effect to removal of errors. In this paper we introduce the completely new and significantly improved version of PDC which was implemented in \textit{Kepler} SOC 8.0. This new PDC version, which utilizes a Bayesian approach for removal of systematics, reliably corrects errors in the light curves while at the same time preserving planet transits and other astrophysically interesting signals. We describe the architecture and the algorithms of this new PDC module, show typical errors encountered in \textit{Kepler} data, and illustrate the corrections using real light curve examples.
\end{abstract}

\newpage \clearpage

\section{Introduction}
\subsection{The \textit{Kepler} Mission}
The \textit{Kepler} Mission is the tenth mission in NASA's Discovery Program. Its main objective is to detect Earth-size planets in the habitable zone\footnote{The habitable zone is the range of orbit radii where temperatures allow for liquid water on the planet's surface.\mycite{kaltenegger11a}} \mycite{kasting93a,koch10a} of Sun-like stars and to determine what fraction $\eta_{\oplus}$ of all stars in the milky way host such planets. Planet detection in \textit{Kepler} is performed using transit photometry~\mycite{deeg98a,charbonneau00a}: A planet transiting between its star and the \textit{Kepler} photometer blocks a fraction of the stars light proportional to the planet's projected area relative to that of its star, and thus decreases the apparent brightness of the star. For that purpose, the \textit{Kepler} spacecraft is in an Earth-trailing orbit and is continuously observing the incoming light of about 156,000 stars in a field of view of 115\,deg$^{2}$. The light flux from the stars is recorded using an array of 42 CCDs, arranged as 84 module output channels each of which has 1100$\times$1024 pixels. Each CCD is read out in a 6.54\,sec interval, and the flux for each target is integrated for 29.4\,minutes, a so-called Long Cadence (LC) interval\footnote{Note that the targets do not have to be stars. Some of \textit{Kepler's} targets are in fact galaxies. However, this distinction is not relevant for the context of this paper.}. The large number of targets is required because the probability for correct geometric alignment of a planet's orbit with the star-to-photometer-axis is the ratio of the star's radius and the planet orbit radius. This translates to a transit probability of only about 0.5\,\% of all Earth-like planets in a 1\,AU orbit around Sun-like stars. Therefore, a sufficiently large number of stars has to be observed to obtain a reliable estimate of the frequency of Earth-like planets in our galaxy ($\eta_{\oplus}$). In addition to these LC targets, a subselection of 512 targets are sampled at 58.35\,sec ``Short-Cadence'' (SC) intervals, which allows not only for more precise transit timing measurement, but also for detailed asteroseismological investigations of these stars, and for measurements of binary star eclipse timing variations.

With the area ratio of the Earth relative to the Sun being $\approx 1e^{-4}$, detection of transits of Earth-like planets requires an unprecedented photometric precision of 20\,ppm~\mycite{koch10a,jenkins10c}. The search for the transits is further complicated by the presence of systematic errors in the photometric data, which occlude the transit signatures as well as other astrophysical signals in the light flux series. The main causes for the systematic errors are changes in the focus of the photometer (caused by instrinsic and extrinsic thermal variations), differential velocity aberration, residual spacecraft pointing errors\footnote{\textit{Kepler's} pointing stability is 0.009'', 3\,$\sigma$ on time scales of 15\,min and greater.}, mechanical vibrations, and electrical interference with other devices onboard the spacecraft (typically leading to Moir\'e like patterns). Here we report on the new Presearch Data Conditioning (PDC) module in the \textit{Kepler} Science Operations Center (SOC) Science Processing Pipeline, which is specifically designed to remove these systematic errors and other artifacts from the light curves prior to planet detection. The new PDC module, which has been completely rewritten for \textit{Kepler} SOC 8.0 (the official \textit{Kepler} software release in August 2011), shows dramatically improved performance -- in particular at removing systematic trends while preserving stellar variability, and at detecting and correcting flux discontinuities.


\subsection{\textit{Kepler} Science Processing Pipeline}
Photometric data from the \textit{Kepler} spacecraft is processed through the \textit{Kepler} Science Processing Pipeline~\mycite{jenkins10a,middour10a} to extract transit information and export light curves that are readily prepared for asteroseismological analysis. The raw pixel data from the photometer are passed to the \textit{Calibration} module (CAL)~\mycite{quintana10a}, which performs pixel-level calibration for target and background pixels. Digital apertures for each target are calculated by the \textit{Target Aperture Definition} (TAD) module~\mycite{bryson10a}. \textit{Photometric Analysis} (PA)~\mycite{twicken10a} then selects the pixels defined by the respective target aperture and extracts the light curves, or flux time series, by adding the flux values of these calibrated pixels. PA also labels anomalies in the data and marks the respective cadences as invalid (``gapped''). The flux output from PA is then passed to \textit{Presearch Data Conditioning} (PDC)~\mycite{twicken10b}, which removes systematic and other errors from the light curves, as described in this paper. These clean light curves from PDC are then passed as input to the \textit{Transiting Planet Search} (TPS)~\mycite{tenenbaum10a,jenkins10b,tenenbaum12a} module, which employs a wavelet-based adaptive matched filter to detect planet transit signatures (``Threshold Crossing Events'', TCEs) in the light curves~\mycite{jenkins02a}. Finally, the \textit{Data Validation} (DV) module~\mycite{wu10a}, performs a series of tests on the TCEs detected in TPS to establish or break confidence in each event, for instance eliminating false-positives resulting from eclipsing binaries, and to extract physical parameters for each system. The results from DV are passed to the \textit{Kepler} Science Team for further analysis and follow-up observations. For storage and publication to the scientific user community, the processed pixels and light curves are exported to the Multimission Archive at STScI (MAST)\footnote{\texttt{http://archive.stsci.edu/index.html}}.

In this manuscript we describe the new Presearch Data Conditioning module in \textit{Kepler} SOC 8.0, which has been completely rewritten and replaces the previous PDC module. We will briefly describe the different kinds of systematic errors typically encountered in the light curves, compare the main systematic error cotrending\footnote{We refer to the detrending here as ``cotrending'' to differentiate it from techniques that do not use prior knowledge of the instrumental signatures, be it from correlation in the light curves or from engineering data.} algorithm of the new module with that of the old module, explain the software architecture of the PDC module, and show real light curve examples of the vastly improved performance of the new PDC module. Since one of the major improvements of this completely new version of PDC is the Bayesian Maximum A Posteriori (MAP) approach to cotrending, we will also refer to the new version PDC in \textit{Kepler} SOC 8.0 as ``PDC-MAP'', and to PDC pre-8.0 as ``PDC-LS'' (where LS stands for ``least-squares''), when comparing the two\footnote{Note however that PDC-MAP and PDC-LS refer to the whole PDC module, rather than only the MAP or LS fitting algorithms.}.


\subsection{Errors in the Light Curves: Tasks of PDC}
\textit{Kepler's} light curves as they come from PA contain a variety of systematic and other errors, which can occlude the astrophysical features in the light curves, and can prevent detection of small planet transits. These errors are caused by a variety of instrumental effects, and span a wide range of frequencies and amplitudes.

The most dominant errors mainly result from Differential Velocity Aberration (DVA), which can introduce low-frequency systematic errors over the whole quarter. Another source of strong systematic errors, typically with periods of hours to days, are focus changes that are caused by thermal transients. Such temperature changes are usually observed during the recovery from planned or unplanned interrupts in the science data collection. Unplanned interrupts include safe-mode events and loss of fine point. Planned events that interrupt science data collection are the monthly ``Earth-points'' and the quarterly rolls. The former is spatial realignment of the \textit{Kepler} spacecraft to point its high-gain antenna towards Earth for data downlink~\mycite{haas10a} and typically lasts about one day. The latter is a 90$^{o}$ rotation around \textit{Kepler's} longitudinal axis every three months to keep the solar arrays directed towards the Sun, and the radiator which cools the CCD away from it~\mycite{haas10a}. Temperature-change related focus changes can also be faster, such as the intermittent modulation of the focus by $\approx 1\,\mu m$ every 3.2 hours by a heater on one of the reaction wheel housings, which was partially mitigated after quarter 1~\mycite{jenkins11a}. The magnitude of temperature-related effects on the focus has been found to be $\approx 2.2 \mu m$ per $^{o}$C~\mycite{jenkins11a}. Fast modes in the systematic errors can also be caused by thermal transients resulting from the reaction-wheel-desaturation on a three day period, and by pointing inaccuracies during the zero-crossings of the reaction-wheels.

\begin{figure} 
\centering
\epsscale{1.0}
\plotone{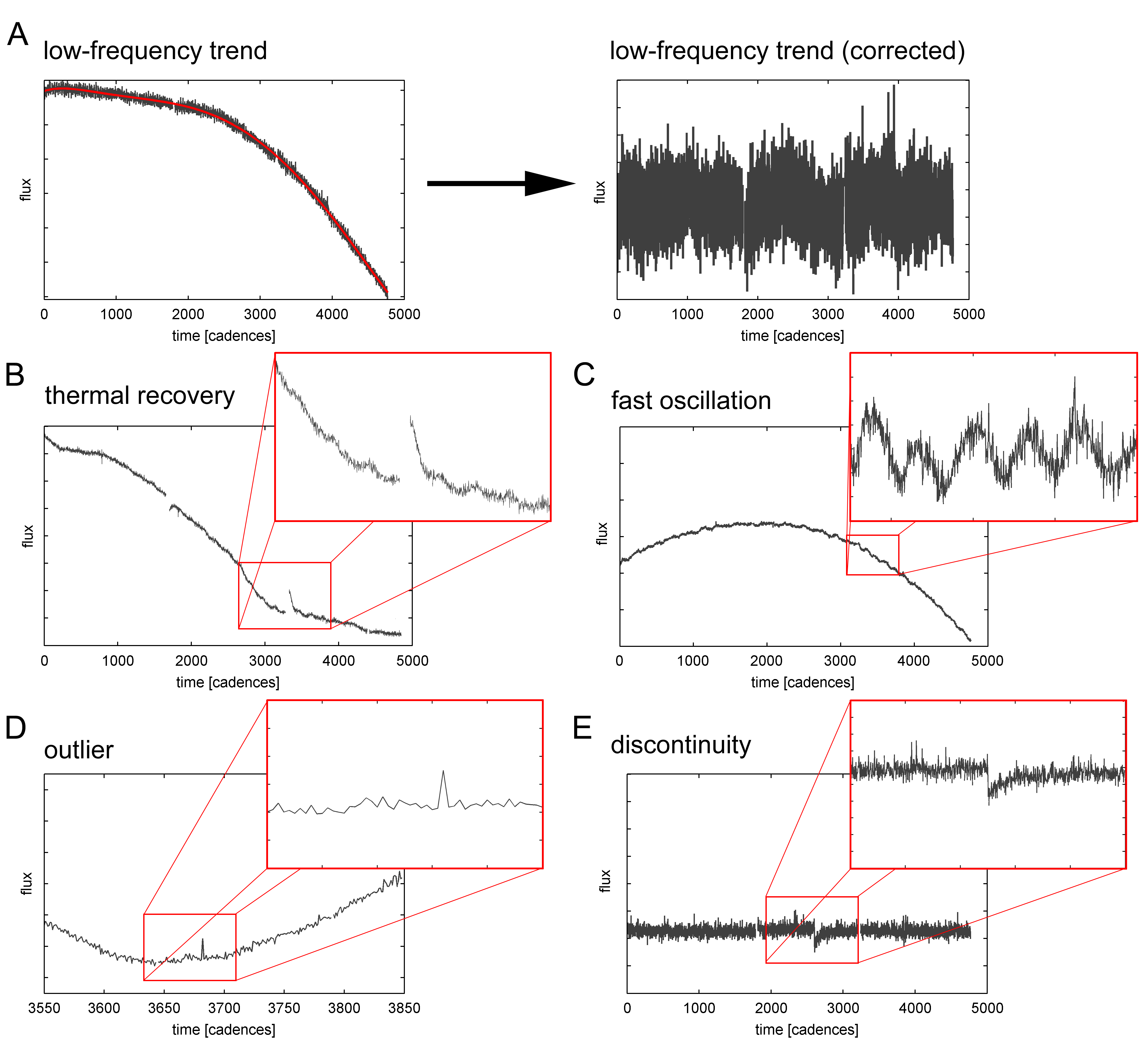}
\caption{\label{fig:systematicerrors}Examples for typical errors in PA light curves, which are corrected in PDC. \textbf{A:} long trend (commonly due to DVA). \textbf{B:} thermal recovery, primarily an exponential recovery after an Earth-point, quarterly roll, or safe-mode. The corresponding gap of approx.\ 50 cadences can be seen right before the recovery. \textbf{C:} fast oscillations, such as from the reaction-wheel desaturation. \textbf{D:} outlier. \textbf{E:} discontinuity (Sudden-Pixel-Sensitivity-Dropoff, SPSD), commonly caused by cosmic rays or solar energetic particles.}
\end{figure}

In addition to these systematic trends and modulations, the PA light curves suffer from local errors such as outliers, discontinuities, Argabrightenings~\mycite{witteborn11a}, and gaps in the data. Further, there are also global corrections that have to be applied to the flux amplitude, such as corrections for incompletely captured target flux (the flux fraction, Section~\ref{sec:FluxFraction}), and for excess light from neighboring stars into the aperture (crowding metric). These effects are illustrated in Figure~\ref{fig:systematicerrors} and described in more detail below.

The task of the PDC module in \textit{Kepler} is to identify and correct all these different kinds of systematic errors, while preserving planet-transits and other astrophysical signals in the light curves.

\section{Architecture and Algorithms of PDC}

\subsection{Cotrending of Systematic Errors in PDC}
The main component of PDC-MAP for removal of systematic trends in the light curves, and arguably the most determinant factor for the significantly improved correction performance as compared to PDC-LS (before \textit{Kepler} SOC 8.0), is the new cotrending routine which is based on a Bayesian Maximum A Posteriori approach~\mycite{degroot70a}. In general, removal of systematic trends in \textit{Kepler} data is based on the assumption that the systematic error component in a light curve has a significant degree of correlation with a set of other time series that can be exploited to identify and remove the systematic error. For PDC-LS, this set of other time series was given by instrument readings on the \textit{Kepler} spacecraft, whereas for PDC-MAP correlations with other light curves are used. We will give a brief review of the PDC-LS cotrending algorithm before describing the new PDC-MAP algorithm in \textit{Kepler} SOC 8.0. As of \textit{Kepler} SOC 8.0, the old PDC-LS cotrending algorithm is still used to process short-cadence data (due to limitations described below), and a logic branch on the controller level calls the respective PDC version for either long-cadence or short-cadence data.


\subsubsection{Cotrending in PDC-LS: Least squares fitting to ancillary engineering data -- \textit{Stellar Variability lost}}\label{sec:CotrendingLS}
Removal of systematic errors in PDC-LS was previously performed based on correlations with a set of ancillary engineering data. These data include the temperatures at the local detector electronics below the CCD array, and polynomials describing the centroid motion of the targets from PA~\mycite{twicken10a}. This approach is based on the assumption that systematic errors in the light curves can be explained by a combination of instrument effects that are expressed in the engineering data, and the motion of target centroids. A robust least squares fit of each light curve to the design matrix constructed from these time series is performed to identify and remove correlated signatures. For further details of the algorithm, see Ref.~\mycite{twicken10b}.

PDC-LS did a good job at correcting light curves, and allowed for many of the plethora of new discoveries in planet detection~\mycite{borucki10a,holman10a,doyle11a,fressin11a,lissauer11a,welsh12a,borucki12a} and stellar astrophysics~\mycite{meibom11a,beck11a,stello11a,chaplin11a} based on \textit{Kepler} data. However, it suffered from two deficiencies for a considerable fraction of light curves.

\begin{figure}
\centering
\epsscale{0.7}
\plotone{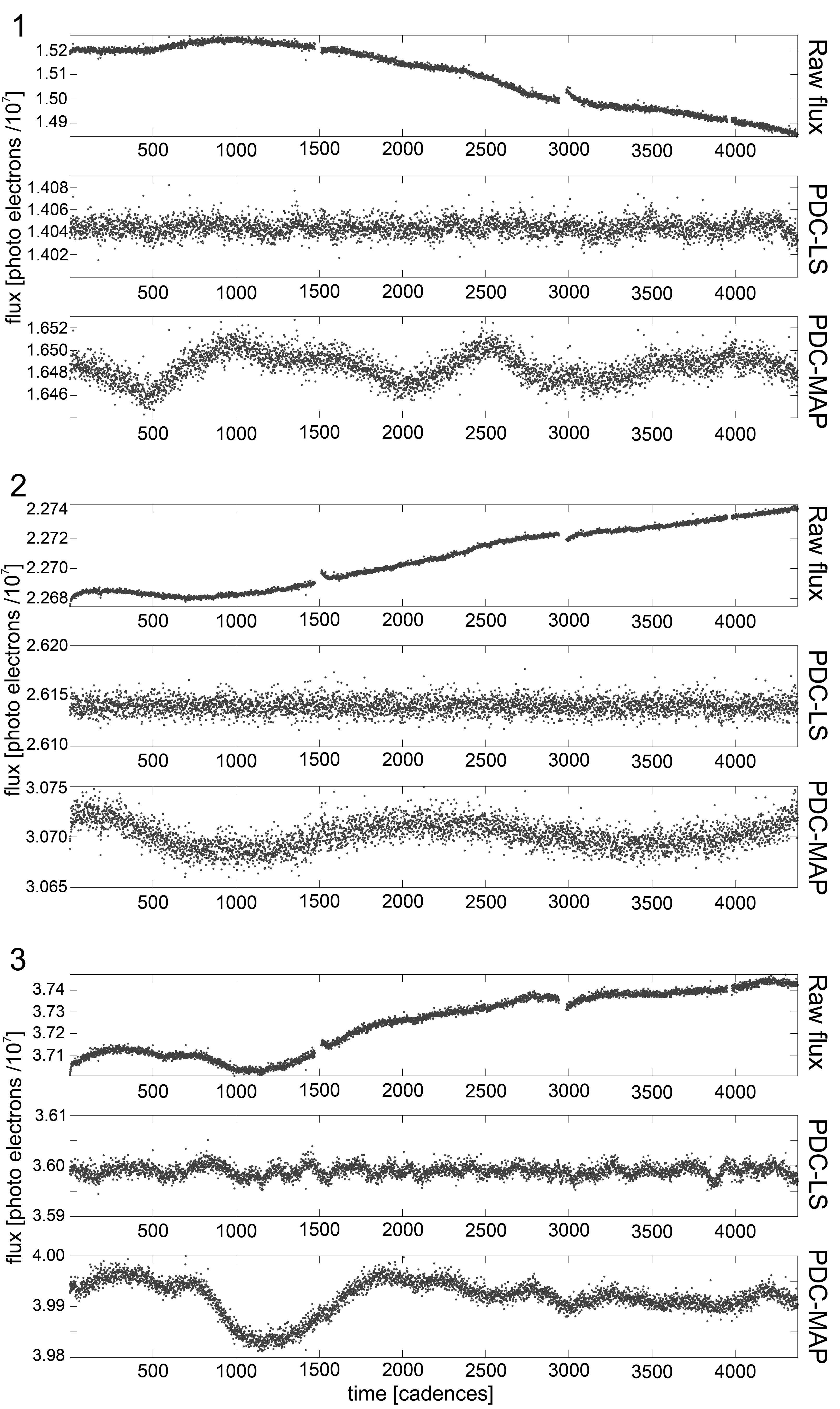}
\caption{\label{fig:overfitting}Examples of light curves where the old PDC-LS overfitted the systematic trends and removed stellar variability. In each of the three panels (1--3), top is the input to PDC (output from PA), middle is PDC-LS (pre \textit{Kepler} SOC 8.0) using least squares cotrending, bottom is PDC-MAP using maximum a posteriori cotrending. Example 3 also shows incomplete cotrending of smaller scale feature in PDC-LS, as can be seen on the residual 3-day reaction wheel cycle signature (see Fig.~\ref{fig:systematicerrors}C).}
\end{figure}

First, the biggest problem observed with PDC-LS was overfitting of the data, which led to removal of stellar variability -- the corrected light curves of many stars appeared overly flat and featureless. This overfitting was due to the least squares approach to cotrending the light curves. With sufficiently many cotrending basis vectors\footnote{The cotrending basis vectors are a set of time series which describe the systematic errors and are used to remove these systematic errors from the light curves by fitting the light curves against the basis vectors (see below).}, coincidental correlations of instrument readings with stellar variability happened frequently, and thus the stellar features in the light curves were mistakenly identified as systematic errors and consequently removed. The Bayesian Maximum A Posteriori approach in PDC-MAP is specifically designed to prevent overfitting and in fact solves this problem very well, as will be shown in this work. See Figure~\ref{fig:overfitting} for some examples of the input light curve to PDC (the uncorrected flux output from PA), the (overfitted) output of PDC-LS, and the (significantly improved) output of PDC-MAP. Note that the PDC-LS and PDC-MAP light curves differ slightly in absolute magnitude, because the flux fraction correction (see Section~\ref{sec:FluxFraction}) was performed differently in PDC-LS.

\begin{figure}[!ht]
\centering
\epsscale{0.5} 
\plotone{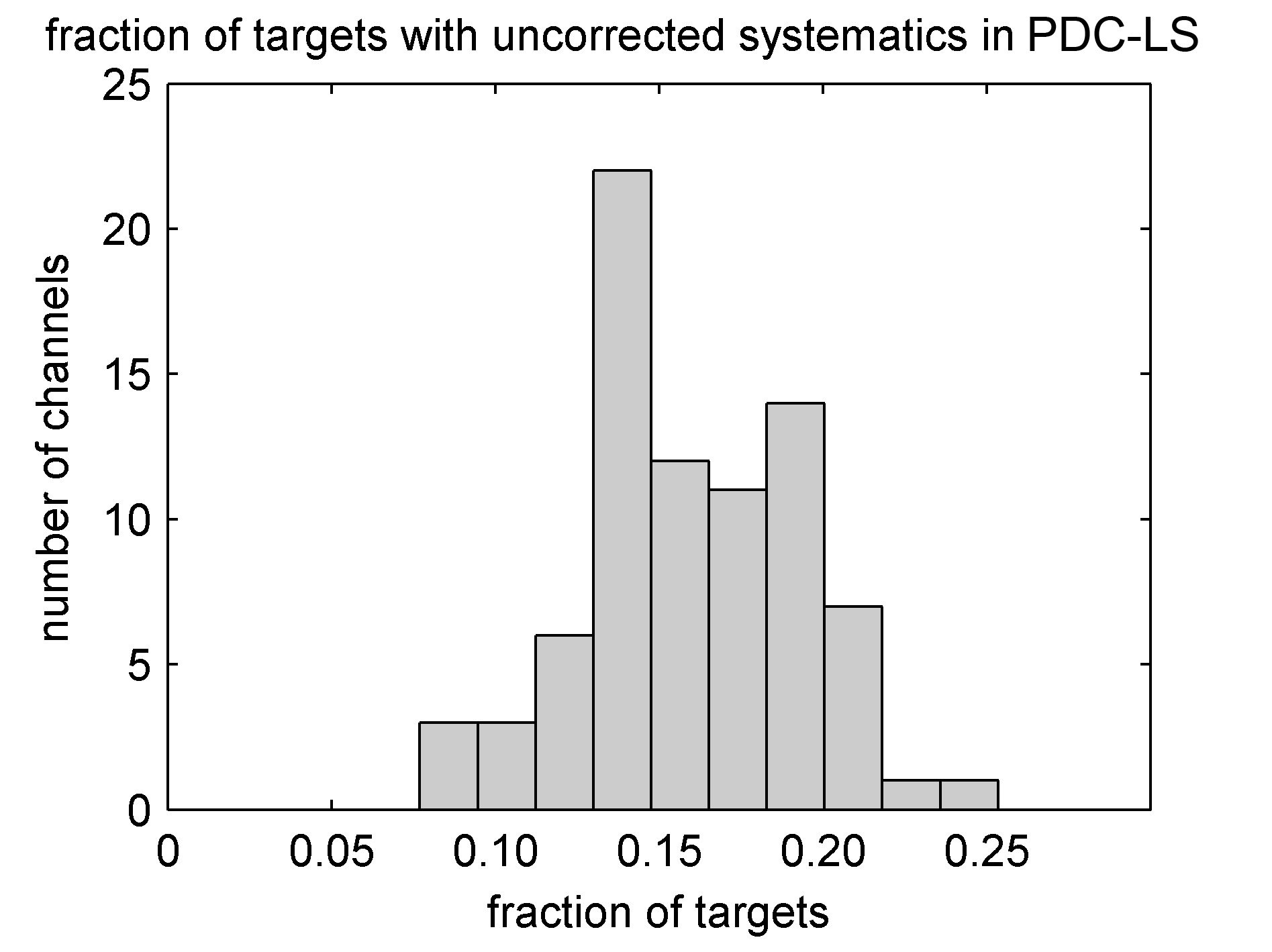}
\caption{\label{fig:hist_pdc7fail_fraction}Fraction of targets per channel for which PDC-LS with least squares fitting to ancillary engineering data failed (quarter 7 data).}
\end{figure}

\begin{figure}
\centering
\epsscale{0.7}
\plotone{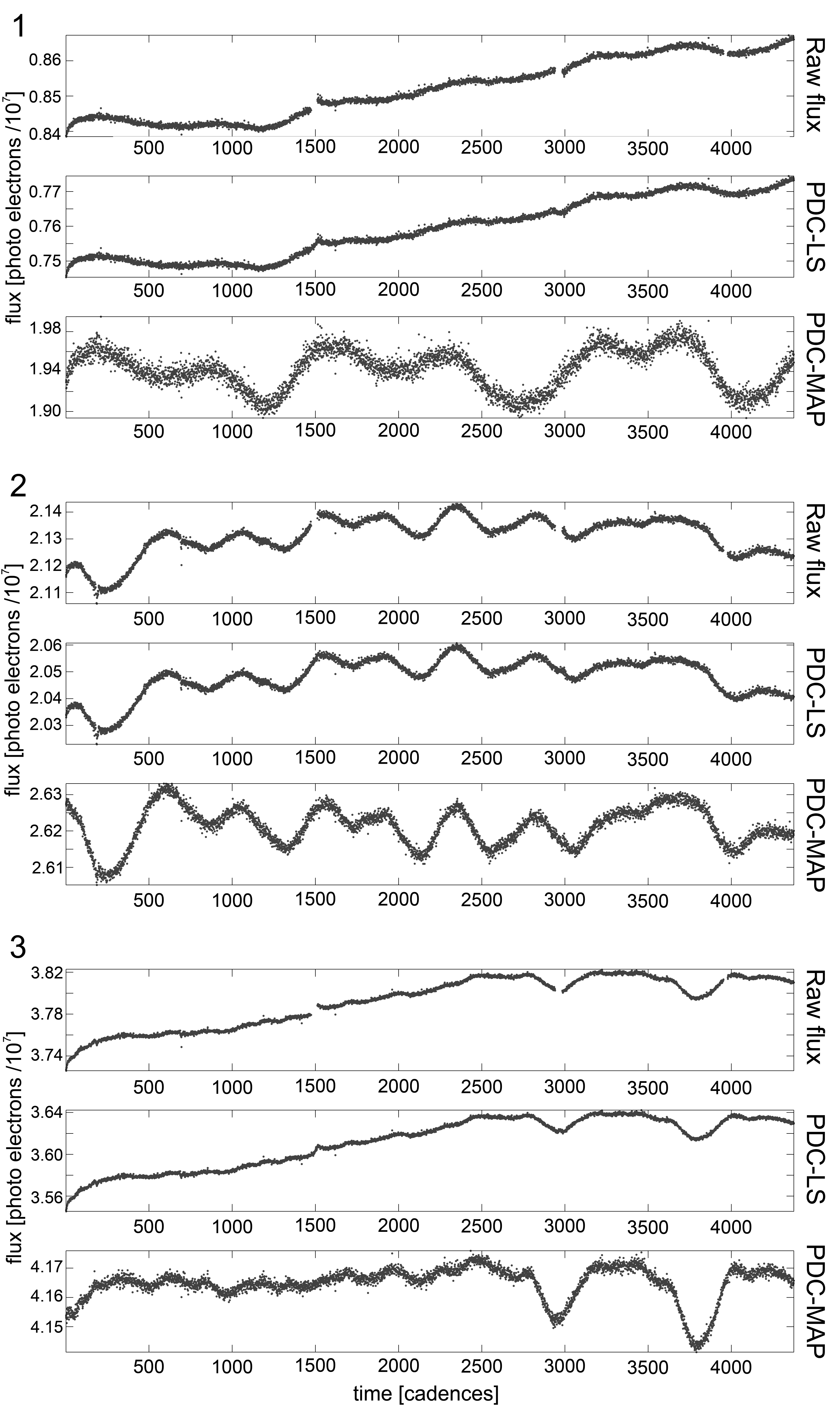}
\caption{\label{fig:pdcgaveup}Examples of light curves where PDC-LS did not perform cotrending because too much noise was injected in the correction. Instead, the uncorrected flux from PA was returned, with only non-trend corrections performed, such as discontinuities, outliers, and flux magnitude.}
\end{figure}

A second problem of PDC-LS was that high-frequency noise was sometimes injected by the correction, as an unwanted side-effect of reducing the bulk root-mean-square deviation. A self-check was in place to prevent too strong noise-injection, in which case PDC-LS rejected the systematic error correction and instead returned a light curve where systematic trends were not removed, but only other corrections (e.g.\ discontinuities, outliers, etc) were applied. The fraction of targets which were not cotrended at all was considerable. For instance, of all 162,926 targets observed with \textit{Kepler} during quarter 7, 25411 (15.6\,\%) light curves were not cotrended. The distribution of non-cotrended targets per channel is displayed in Figure~\ref{fig:hist_pdc7fail_fraction} and ranges from 7.7\,\% to 25.4\,\% per channel. Figure~\ref{fig:pdcgaveup} shows some examples of light curves that could not be corrected properly with PDC-LS. The problem of noise-injection has been significantly reduced in PDC-MAP, as can be seen in the bottom panels of these examples.

\subsubsection{Cotrending in PDC-MAP: Bayesian Maximum A Posteriori fitting to the ensemble of light curves -- \textit{Stellar Variability regained}}\label{sec:CotrendingMAP}
The new version of PDC, PDC-MAP, uses a different approach in mainly two regards. First, instead of employing ancillary engineering data, cotrending is performed against the ensemble of light curves on the same channel. Secondly, rather than using a simple least squares fit, a Bayesian Maximum A Posteriori (MAP) approach is used for fitting. Note that while this approach does not use any engineering data explicitly, it is still based on the assumption that instrumental signatures are the main cause for systematic errors in the data. Moreover, it assumes that these signatures are highly correlated between targets that are in proximity on the same CCD channel. This second assumption is a critical ingredient for the Bayesian MAP fit, as it allows for the generation of constraints for the fit by using information about systematics in targets within a neighborhood of the target under investigation. It is mainly these constraints that prevent overfitting, and thus help to preserve stellar variability. The MAP algorithm in PDC is described in detail in a companion paper~\mycite{smith12a}. Here we will only provide a brief overview to an extent that is relevant in the context of this paper.

Cotrending is performed by fitting each light curve to a set of basis vectors, and constraining the fit coefficients in a Bayesian Maximum A Posteriori way by using prior information. The basis vectors are generated from the 50\,\% most correlated target light curves on the channel, in order to have only the strongest correlated trends in the basis vectors and largely exclude targets with substantial individual fluctuations due to stellar variability. Using Singular Value Decomposition (SVD) on this set of light curves, the N largest singular vectors are used as basis vectors. The number of basis vectors to use can be specified as an input parameter to PDC, with eight basis vectors being the default\footnote{One potential future improvement for PDC is to calculate the optimum number based on properties of the singular values and vectors, and potentially iterating using the Goodness Metric (see below).}. By construction, these N basis vectors represent the majority of the systematic errors contained in the light curves. Each of the light curves is projected onto these N basis vectors using a least squares fit. To make this fit more robust and prevent overfitting, however, the fit is constrained using prior information from targets which are expected to be affected by similar systematic errors. For that purpose, a target neighbor list is generated for each target with respect to their distance in a 3D space spanned by stellar magnitude (Kp), Right Ascension (RA), and Declination (Dec). The latter two coordinates, RA and Dec, directly translate into the spatial proximity of two targets on the CCD. Further, Kp is used as an additional dimension because systematic errors are not only correlated with location on the CCD but also with signal amplitude.

The contribution of each basis vector $j$ to each light curve $i$ is described in the fit coefficient $\Theta^{i}_{j}$. However, instead of simply using the fit coefficient of the least squares fit, the most likely coefficient given the observations in the neighborhood of the target is picked. In Bayesian terminology, the posterior probability density function (PDF) $p(\Theta|y)$ is maximized, given a prior PDF $p(\Theta)$ generated from fit coefficients of all neighboring targets, weighted by their distance to target $i$:
\begin{equation}
 p(\Theta|y) = \frac{p(y|\Theta) \cdot p(\Theta)}{p(y)} = \frac{p(y|\Theta) \cdot p(\Theta)}{\int p(y|\Theta)p(\Theta)d\Theta}
\end{equation}
The $\Theta$ that maximizes $p(\Theta|y)$ is used as the fit coefficient for the MAP fit. Note that the denominator $p(y)$ is simply a normalization over all possible observations $y$ and can in practice be omitted in the maximization process. Taking the maximum a posteriori $\Theta$ effectively constrains the fit and helps to avoid overfitting. In particular, only light curve features that are also present in targets in the neighborhood will be removed, whereas any stellar variability that might have coincidental correlation with some of the basis vectors will be constrained by the prior, and hence be preserved.

\begin{figure} 
\centering
\epsscale{0.95}
\plotone{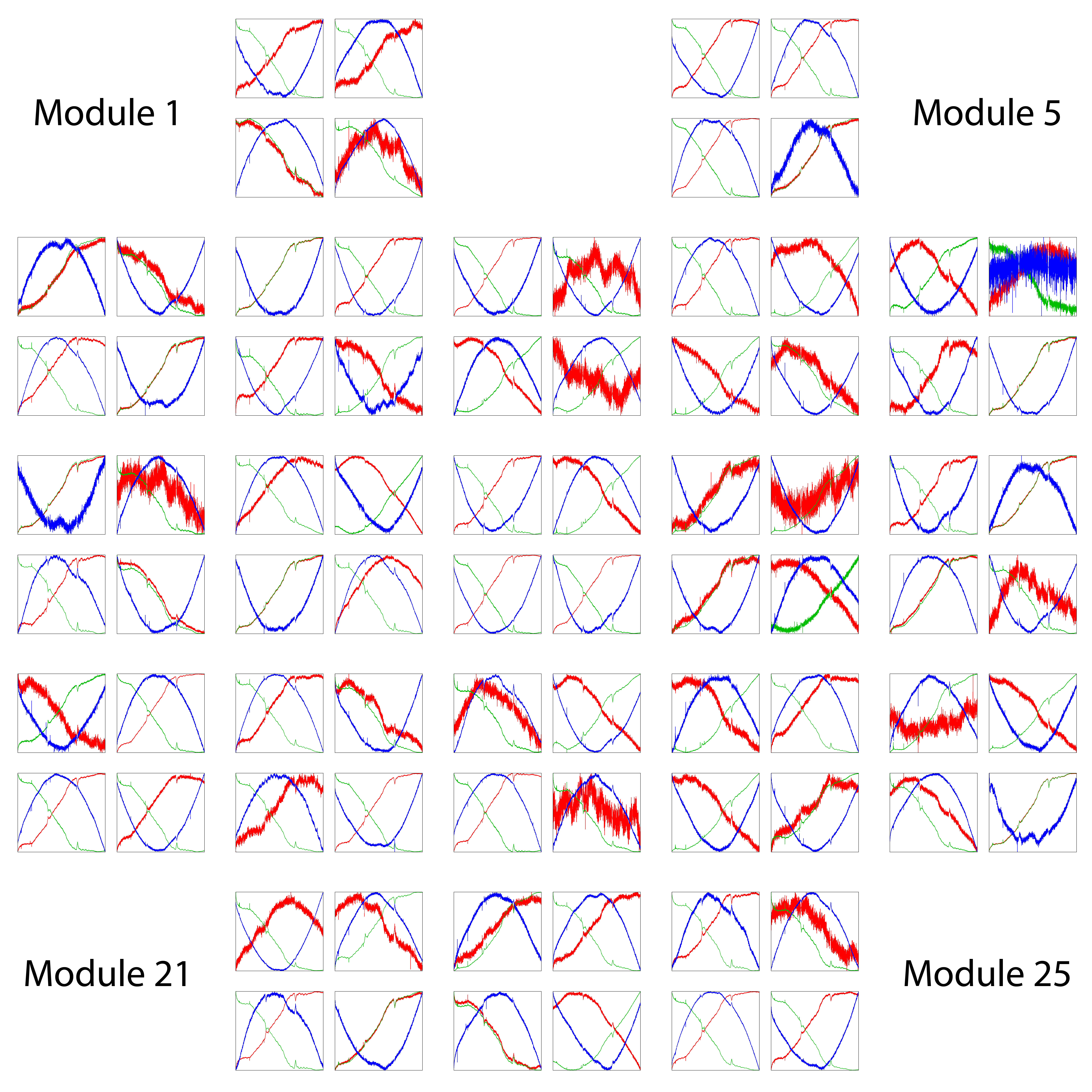}
\caption{\label{fig:basisvectorsccdq7}First three cotrending basis vectors for each channel on the \textit{Kepler} CCD, data from quarter 7. Note that the four channels of Module 3 are missing because that module failed during quarter 4.}
\end{figure}

Figure~\ref{fig:basisvectorsccdq7} shows the first three basis vectors for each channel on the CCD for quarter 7. This illustrates how the dominant systematic errors vary across the field of view of the CCD. However, some trends are observed on almost all channels.

\subsection{Inputs to PDC} 
\begin{figure}[!ht]
\centering
\epsscale{0.7}
\plotone{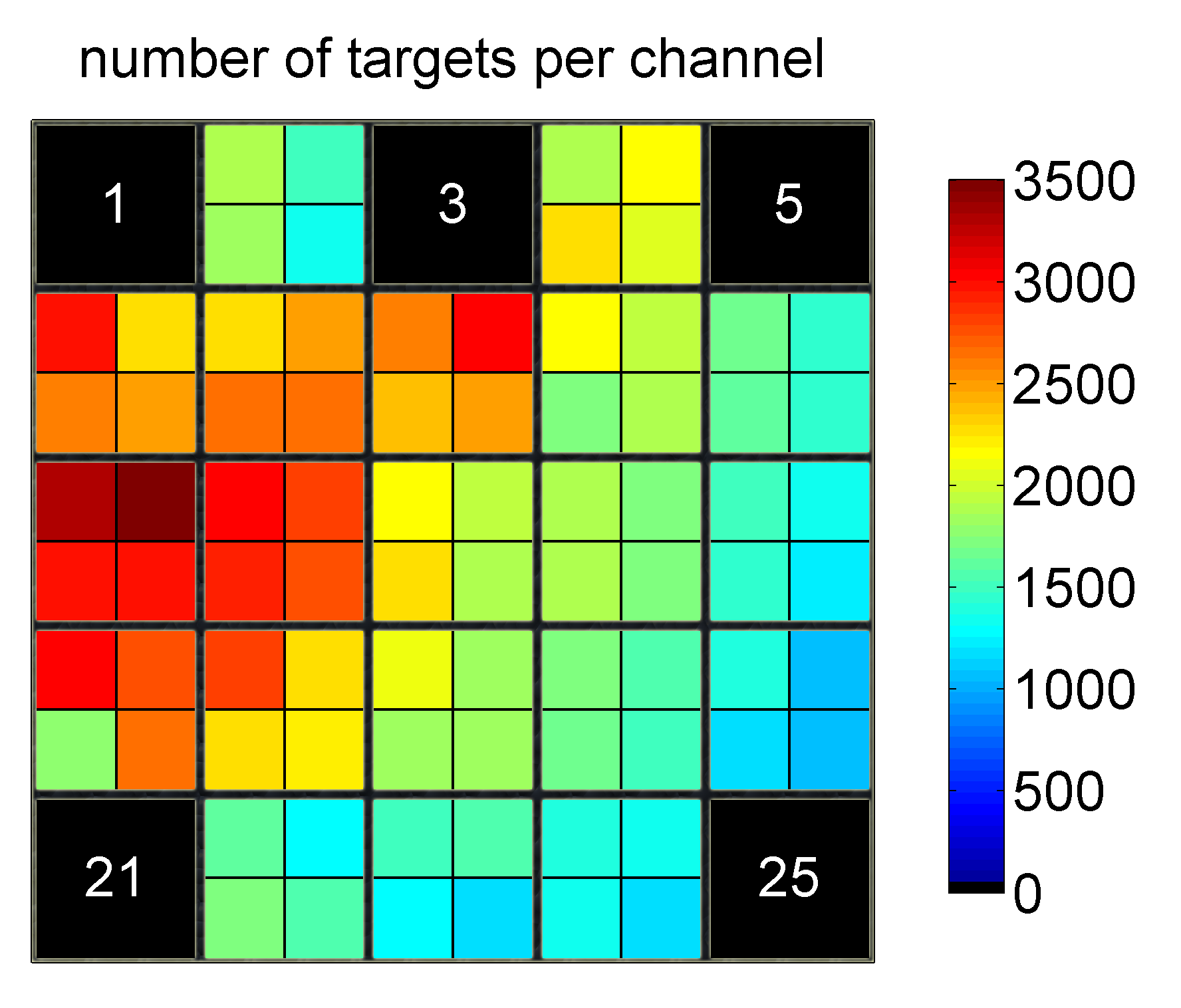}
\caption{\label{fig:CCDtargets}Number of targets per channel in quarter 7. Each of the 25 modules is divided into four module outputs (channels). Modules 1, 5, 21, and 25 are used for pointing and calibration purposes and not for collection of science data. Module 3 failed in quarter 4, leaving a total of 80 channels. There are significantly more targets in the upper left corner because that section of the \textit{Kepler} field of view is closer to the galactic plane.}
\end{figure}

The inputs to PDC are the calibrated, uncorrected flux time series (measured in photoelectrons), as prepared by PA. One unit of work for PDC is a single module output (``channel'') from the CCD with the duration of three months, with a sampling period of 29.4\,minutes. Thus, there are typically 1000--3500 targets per channel (see Fig.~\ref{fig:CCDtargets}), and about 4500 data points (``cadences'') per target\footnote{This is for ``long cadence'' data, which is addressed here. ``Short cadence'' data is sampled at a rate of 57.8\,seconds for the duration of one month, and there are only 512 targets in total on all 84 module outputs.}. In addition to the flux time series, the target data structure also contains per-cadence flux uncertainty values and gap indicators to denote invalid cadences, and per-target data such as the flux fraction and crowding metric (see below), as well as bookkeeping information such as the \textit{Kepler} ID and KIC\footnote{The \textit{Kepler} Input-Catalog (KIC) is the primary source of information about targets observed with \textit{Kepler}, and contains their RA/Dec coordinates, estimates for a star's radius, temperature, surface gravity, and other data. URL: \texttt{http://archive.stsci.edu/kepler/kepler\_fov/search.php}} data for each target.

\begin{figure}[!ht] 
\centering
\epsscale{1.0}
\plotone{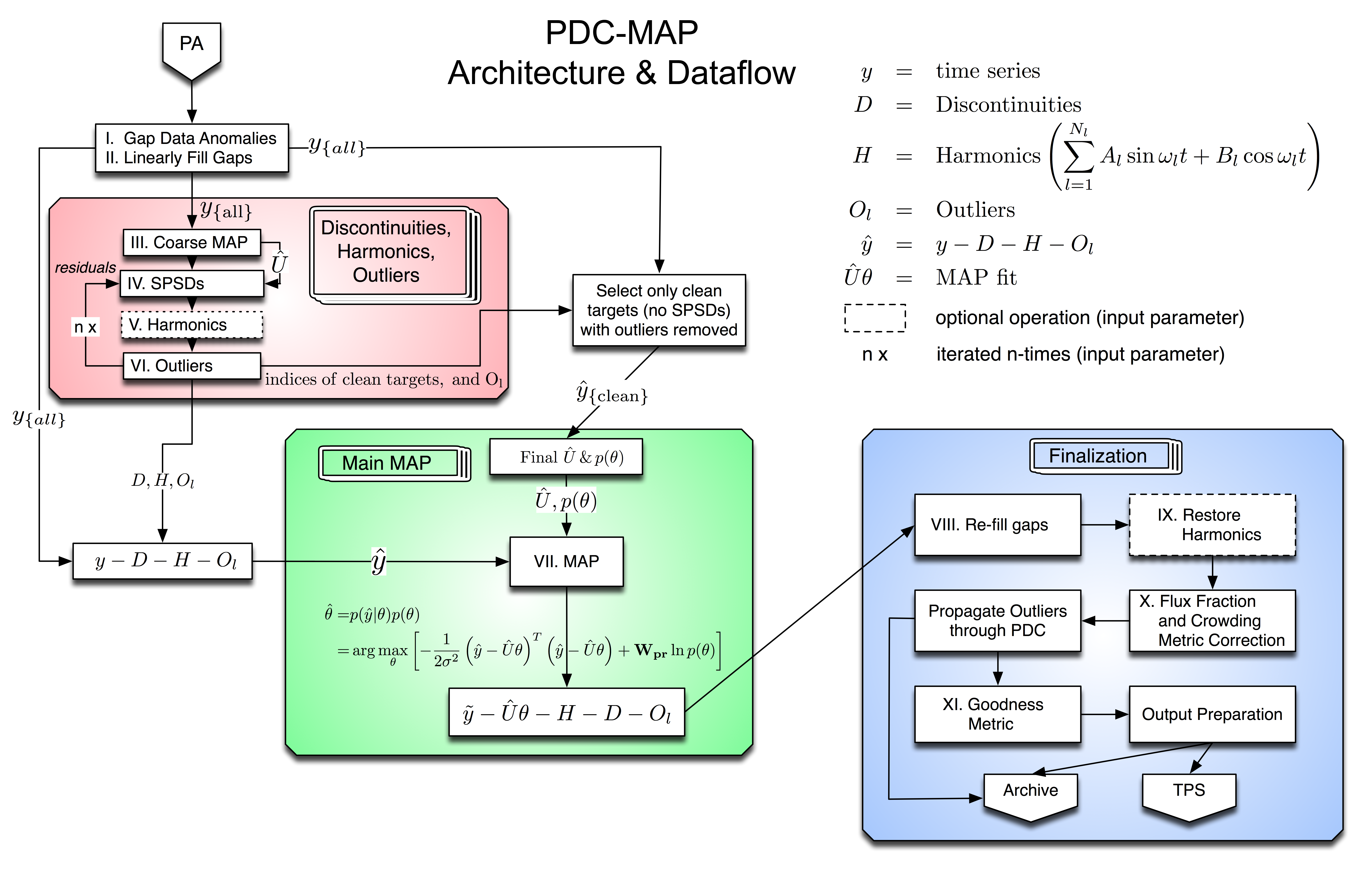}
\caption{\label{fig:PDC_flow_chart}Architecture and Data Flow of PDC-MAP as of \textit{Kepler} SOC 8.0.}
\end{figure}


\subsection{Overview and Data Flow}\label{sec:DataFlow} 
\begin{figure}[!ht] 
\centering
\epsscale{0.8} 
\plotone{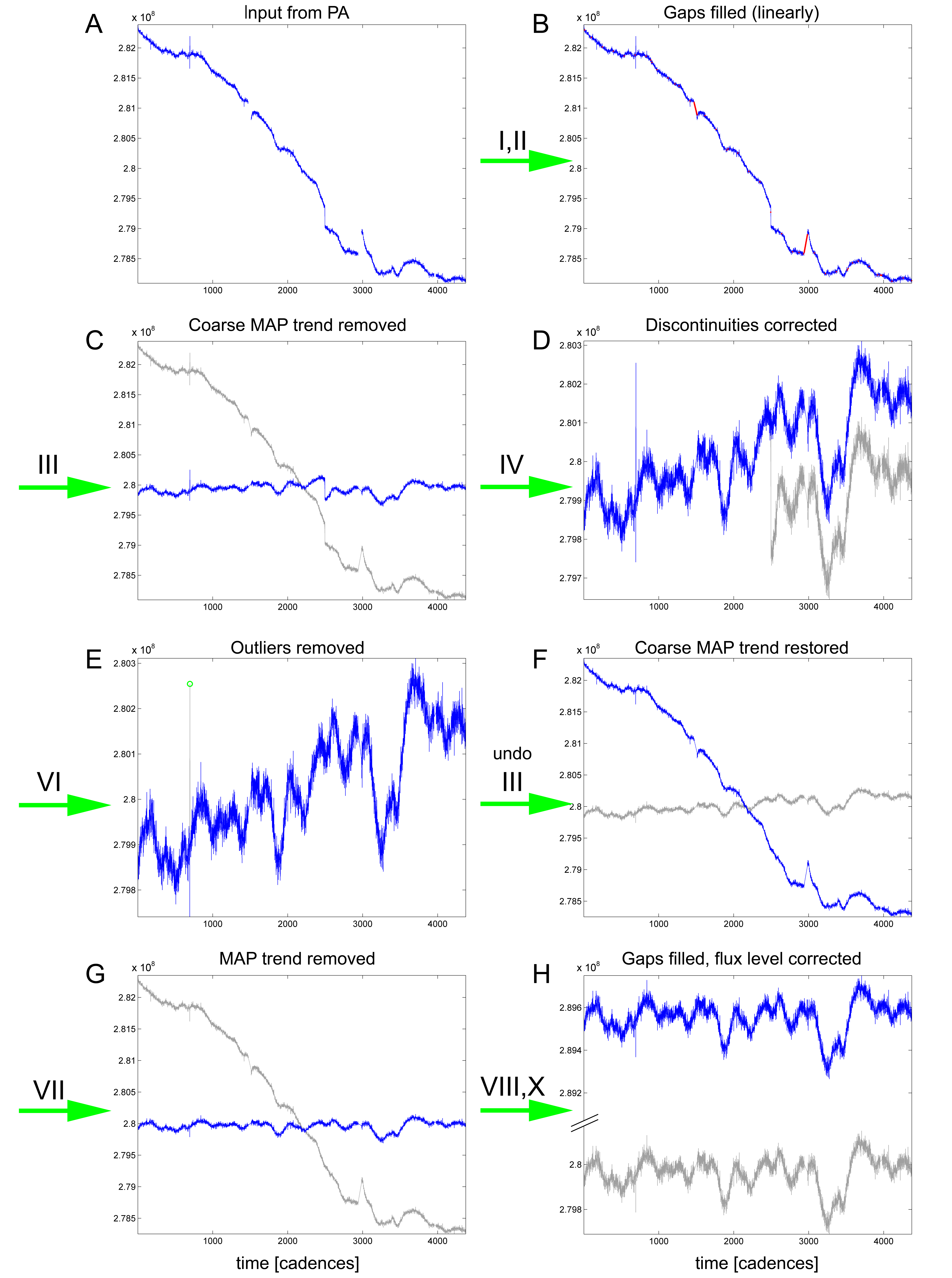}
\caption{\label{fig:target850walkthrough}\small Example for light curve processing of a variable star in PDC, from the output of PA and input to PDC (top right), to the output of PDC (bottom right). The numbers above the arrows denote the operations as defined in Section~\ref{sec:DataFlow} to get from the previous state (which is shown again in gray for comparison) to the current state (blue curve). \textbf{A:} Input to PDC. \textbf{B:} Gaps have been linearly filled. \textbf{C:} The coarse MAP trend has been removed, this is a temporary correction to facilitate discontinuity correction and outlier removal. \textbf{D:} Discontinuities have been corrected. \textbf{E:} Outliers have been removed. \textbf{F:} The coarse MAP trend has been restored. \textbf{G:} The final MAP fit has been removed. \textbf{H:} Gaps have been filled using a more sophisticated algorithm, and the absolute flux magnitude has been adjusted by flux fraction and crowding metric corrections.\normalsize}
\end{figure}

The data flow in PDC-MAP is displayed in Figure~\ref{fig:PDC_flow_chart}. Figures \ref{fig:target158walkthrough} and \ref{fig:target850walkthrough} show two light curve examples as they are being processed throgh PDC. The individual operations of this sequential processing are described below:

\begin{figure}[!ht] 
\centering
\epsscale{0.8} 
\plotone{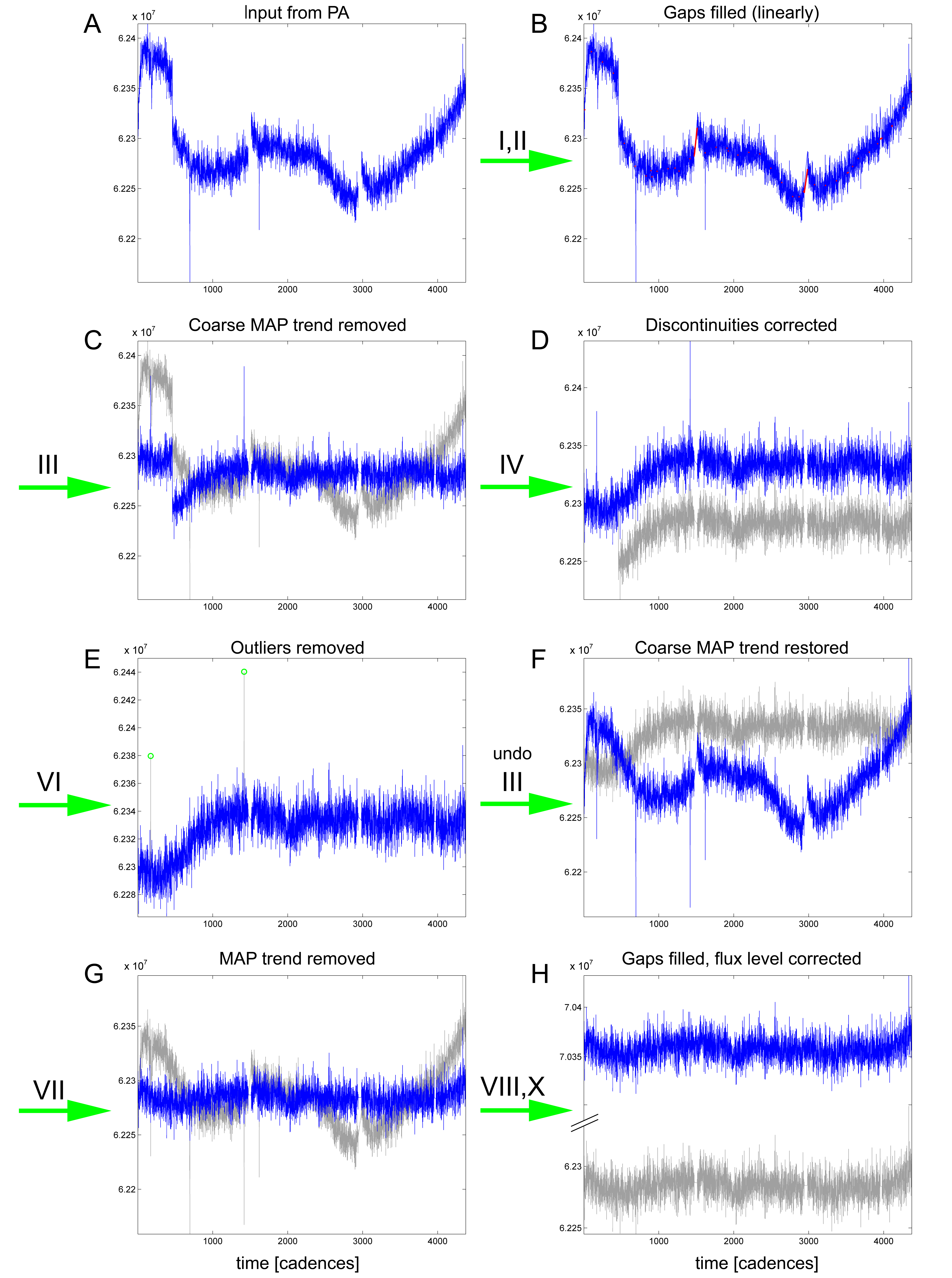}
\caption{\label{fig:target158walkthrough}\small Example for light curve processing of a quiet star in PDC, from the output of PA and input to PDC (top right), to the output of PDC (bottom right). The numbers above the arrows denote the operations as defined in Section~\ref{sec:DataFlow} to get from the previous state (which is shown again in gray for comparison) to the current state (blue curve). \textbf{A:} Input to PDC. \textbf{B:} Gaps have been linearly filled. \textbf{C:} The coarse MAP trend has been removed, this is a temporary correction to facilitate discontinuity correction and outlier removal. \textbf{D:} Discontinuities have been corrected. \textbf{E:} Outliers have been removed. \textbf{F:} The coarse MAP trend has been restored. \textbf{G:} The final MAP fit has been removed. \textbf{H:} Gaps have been filled using a more sophisticated algorithm, and the absolute flux magnitude has been adjusted by flux fraction and crowding metric corrections.\normalsize}
\end{figure}

\subsubsection{I. Gap Data Anomalies} The first step is to flag anomalies in the flux series and mark the respective cadences as not usable (``gapped''). Anomalies may affect only one target (e.g.\ Undershoot Columns~\mycite{bryson10a}) or all targets on the channel (e.g.\ monthly Earth Point downlinks, attitude adjustments, spacecraft Safe-Modes, loss of fine-pointing, and Argabrightenings~\mycite{witteborn11a}), and can have a duration of one to several hundred cadences. The information about the gaps and the type of anomaly is provided separately by PA. Gapped data is labeled as such and is not exported to the MAST.


\subsubsection{II. Linearly Fill Gaps} Even though gapped data is not being exported, some operations in PDC require contiguous data over the whole time series. Therefore, gaps are filled by linear interpolation for internal use. Note that these linearly interpolated gap values are filled a second time at the end of PDC, using a more sophisticated algorithm (see below) to prepare the light curves for TPS.


\subsubsection{III. Coarse Cotrending} A first coarse cotrending is performed, using the Bayesian Maximum A Posteriori (MAP) approach described above (\ref{sec:CotrendingMAP}). This coarse cotrending is only applied temporarily to the flux series to enhance the performance of the next three steps (IV, V, VI), and the coarse correction is removed afterwards in order to perform the proper MAP correction (step VII).


\subsubsection{IV. Sudden Pixel Sensitivity Dropoff (SPSD) Correction} Approximately 3\,\% of the light curves contain one or more noticeable downward step discontinuities per quarter. The vast majority of these discontinuities are the result of cosmic rays or solar energetic particles striking the photometer and causing permanent local changes in pixel sensitivity, although partial exponential recovery is often observed~\mycite{jenkins10a}. Depending on the incident angle and the energy of the cosmic ray, up to a few pixels can be affected, and the net decrease in quantum efficiency is typically $\approx$0.5\,\% for a light curve. Since these SPSDs are not correlated systematics, they can not be removed by the MAP algorithm (see step VII), and instead a specific submodule exists to detect and correct SPSDs. The Discontinuity Correction algorithm used in PDC-LS had a considerable false-negative rate, and was moreover performing a rather crude correction which did not take partial recoveries into account. The new SPSD module in PDC-MAP is significantly more sophisticated, and performs markedly better at detecting and correcting SPSDs. The details of this new algorithm are described in a separate publication~\mycite{kolodziejczak12a}. In brief, SPSD detection is performed by fitting a window of the data to a combination of basis functions, with one of the basis functions modeling a step discontinuity in the flux. The fit coefficient of this basis function is evaluated in the context of flux variability of the light curve ensemble on that channel, to determine the probability for an underlying SPSD. Further checks are performed to prevent false positive detection of transit-like features as discontinuities. Correction then proceeds as a two-stage process, with the first stage estimating the step size (based on analysis of the entire flux series), and the second stage modeling the recovery process. The sum of these two components is the total correction applied to the light curve.

\begin{figure}[!ht]
\centering
\epsscale{0.5}
\plotone{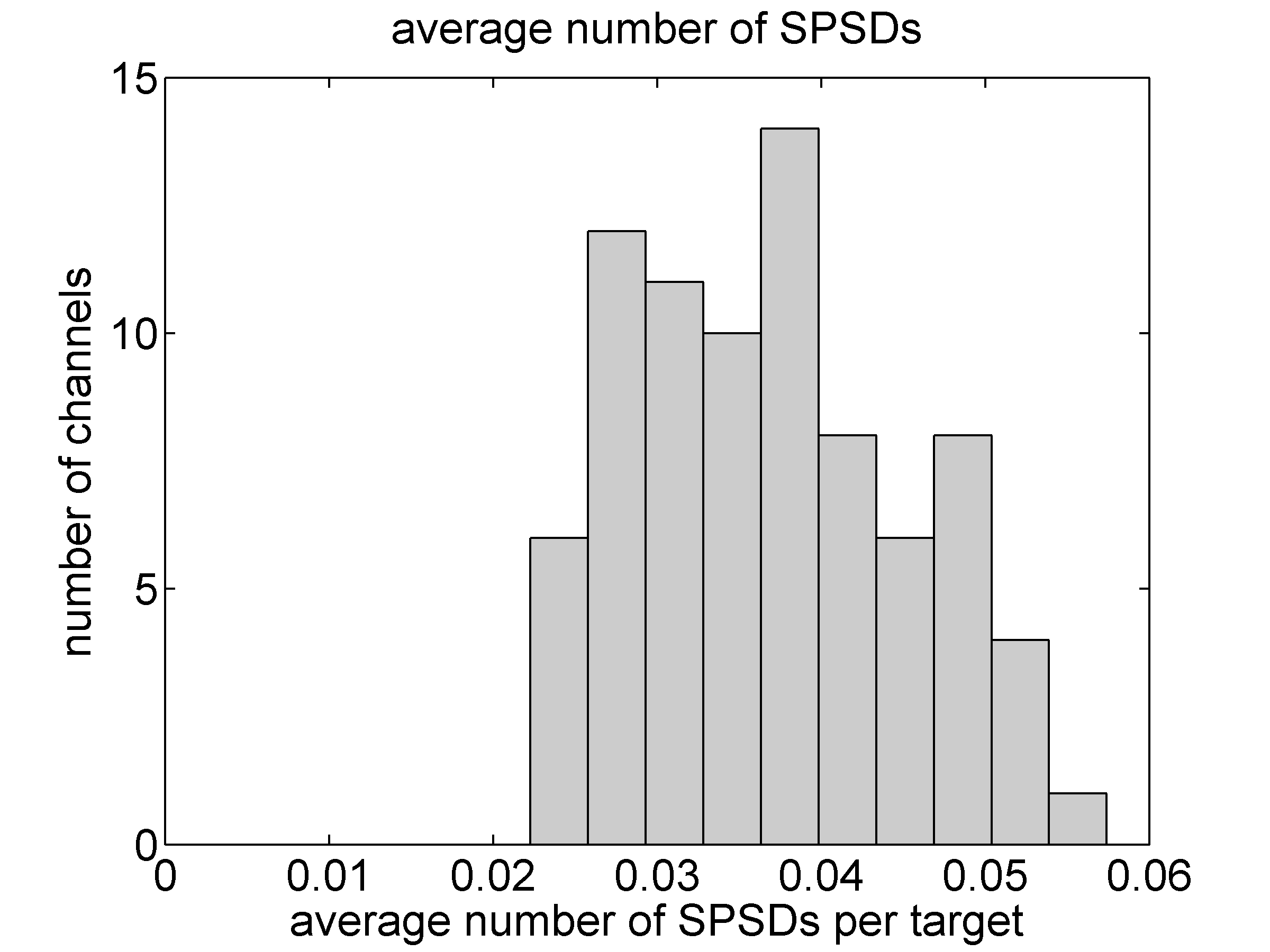}
\caption{\label{fig:hist_spsd_fraction}Fraction of targets per channel for which an SPSD has been detected in Quarter 9. The total number of channels is 80, and not 84, because module number 3 failed in Quarter 4, reducing the number of data channels by four.}
\end{figure}

In the Quarter 9 data, SPSDs were detected in 5252 out of 167404 targets (3.1\,\%). Figure~\ref{fig:hist_spsd_fraction} shows a histogram of the fraction of targets containing SPSDs.



\subsubsection{V. Harmonics Removal (\textit{optional})} Harmonic content can be identified and removed from the light curves of harmonically variable targets. This step was a mandatory operation in PDC-LS in order to prevent harmonic content from corrupting the least squares fitting in the cotrending routine. In PDC-MAP the maximum a posteriori fitting approach is robust against uncorrelated target-specific harmonics and this step is therefore not performed per default, but can be requested with an input parameter. If harmonics are removed for further processing in PDC they are restored later after cotrending. Both corrected harmonics-free as well as corrected light curves with harmonics are exported.

From a theoretical point of view, one might expect that this (previously required) operation of Harmonics Removal prior to cotrending would improve the cotrending -- simply because it reduces the feature complexity of the light curves. However, we found that it has virtually no effect with the new MAP cotrending method. In a direct comparison of 1075 light curves processed with and without Harmonics Removal, we found no difference at all for 97\,\% of the targets. For the remaining 3\,\%, performing Harmonics Removal before cotrending resulted in weak residual long term trends, which had been identified as very low frequency harmonics by the Harmonics Detector. Thus, by removing them before cotrending, and restoring them afterwards, these long term trends were preserved in the light curves. Without knowing the ground-truth, one obviously can not know with certainty whether these long term trends are actually astrophysical effects (in which case they should be preserved), or systematics errors (in which case they should be removed). However, for most of these cases the low frequency trends seemed to be systematic errors, as judged by comparison with the systematic trends identified for other targets, and therefore performing Harmonics Removal appeared to actually decrease the quality of the error correction.

In summary, while Harmonics Removal was a required step in the least squares based cotrending in PDC-LS, it seems that this operation does not improve the error correction (and may even do harm in a few cases) in the new Bayesian approach of PDC-MAP. Therefore, this step is skipped by default.


\subsubsection{VI. Outlier Correction}
Outliers in each flux series are identified based on the mean and standard deviation in a sliding window. The window size and outlier detection threshold are input parameters to PDC. Only isolated (single cadence) points above the threshold are considered to be outliers in order to prevent flagging of planet transits or stellar flares as outliers. These outlier values are ``corrected'' by replacing them with linearly interpolated values. The purpose of this correction is to prevent false triggering of Threshold-Crossing-Events (TCEs) in TPS and also to improve the performance of other operations downstream in PDC. Before export to the MAST, the original outlier values and uncertainties are restored to the corrected light curves and propagated through the PDC correction steps like all regular data points.


\clearpage

\subsubsection{Iteration of SPSD Correction, Harmonics Removal, Outlier Correction} The operations of SPSD Correction (IV), Harmonics Removal (V), and Outlier Correction (VI) are non-orthogonal to each other. In theory, performing them simultaneously rather than sequentially should yield optimal correction, however at the expense of a substantially more complex algorithm that has not yet developed. Iteration of these operations is an approximation to such a joint fitting approach, and an input parameter was included to iterate the operations IV--VI for improved correction performance. In particular, at the time of the design of the architecture it was expected that Harmonics Removal would improve the performance of the SPSD correction~\mycite{kolodziejczak12a} (step IV), and hence that iterating operations IV--VI would be beneficial for the overall correction. However, it was found that the presence of harmonics in the frequency- and amplitude-range that is observed in the \textit{Kepler} data does not decrease the detection performance of the SPSD module by more than 1\,\%, while removing them would come with the disadvantages mentioned above for step V. For these reasons, the operations IV--VI are per default only performed once in PDC-MAP, while leaving the option to perform an iteration in case future design changes should make this desirable.


\subsubsection{VII. Bayesian MAP approach for cotrending to remove systematic errors}
This operation constitutes the main step for the correction of systematic trends in the light curves. Each flux time series is cotrended against the ensemble of light curves in order to remove correlated trends using a Bayesian Maximum A Posteriori approach as described above (\ref{sec:CotrendingMAP}). Before cotrending, the correction applied by the Coarse Cotrending (step III) is restored. The main cotrending procedure (this step) is the same as the routine performed in the Coarse Cotrending (step III), but with two important differences that make its correction significantly more accurate. First, only ``clean'' targets, for which no discontinuities had been found (step IV), are considered to generate the basis vectors. Secondly, discontinuities, harmonics (if enabled), and outliers have been removed from the light curves before the fit, which improves the quality of the correction.


\subsubsection{VIII. Fill Gaps for TPS} Transiting Planet Search (TPS) requires contiguous data points for all cadences in a light curve, and moreover assumes statistical continuity of the wavelet coefficient variances of the flux time series. The simple linear gap filling operation does not meet this requirement. Therefore, a more sophisticated gap filling algorithm is employed to refill the gaps of the corrected light curves before export to TPS. The gap filler internally distinguishes between gaps with ``short`` and with ``long`` duration. The boundary for short and long gaps is determined by the gap filling module parameter set, and typically about 125 cadences. Gaps shorter than this threshold are filled with an autoregressive algorithm that estimates sample values in the gaps with a linear prediction based on the correlation in the neighborhood of the gap. For gaps longer than this threshold, samples away from the edges would tend to become white noise samples, and a different algorithm is therefore used to estimate the gap values: Segments on either side of the gap are reflected onto the gap, and a weighted sum of the two segments is taken where the weighting factor for each segment is decreasing linearly with the distance to respective edge of the gap. This procedure preserves statistical continuity of the data, and in particular, the correlation structure. However, some smoothing at small scales occurs due to the averaging of two noisy signals. To compensate for this effect, the variances of the wavelet coefficients of the filled gap are adjusted to match those of the original data in the neighborhood of the gap. Preserving the wavelet coefficient statistics is of particular relevance because TPS uses a wavelet-based approach for transit detection~\mycite{jenkins10b,jenkins02a,tenenbaum10a}.


\subsubsection{IX. Restore Harmonics (\textit{optional})} If harmonic content had been removed from the light curves (optional step V), then it will be added back here, properly cotrended with MAP. Harmonic-free corrected light curves are exported in addition to the original corrected light curves. Per default, Harmonics Removal is not performed in PDC-MAP as of \textit{Kepler} SOC 8.0 (see step V).


\subsubsection{X. Crowding Metric and Flux Fraction Correction}\label{sec:FluxFraction} When generating flux series from pixel data in PA, the optimal aperture~\mycite{twicken10a} for each target may include additional light from nearby light sources. If not removed, this excess flux would decrease the apparent planet transit depth and lead to a systematic underestimation of planet radii. The so-called \textit{crowding metric} for each target is computed in TAD, and reflects which fraction of the flux in the optimal aperture is due to the target itself. It is a scalar value representing the average over the effective date range. Note that using a scalar average for the crowding metric is an approximation since background sources of light may enter or exit the optimal aperture over the course of the three-month data period. Similarly, using a scalar value for the flux fraction is an approximation because the centroid of the point-spread-function of a target can move over the course of a quarter\footnote{The maximal motion of a target is $\approx$0.6\,pixel over 3 months.}.

Similar to excess flux leaking into the optimal aperture, a fraction of the point-spread-function (PSF) of the target may not be captured in its optimal aperture. To account for this missing fraction, the \textit{flux fraction}, computed by PA, is used to normalize the flux.

Together, these two corrections can be expressed as:
\begin{equation}
\tilde{y}(t) = \frac{ y(t) - (1-c) \cdot {\rm median}(y(t)) }{f} \quad ,
\end{equation}
where $\tilde{y}(t)$ is the corrected cotrended flux, $y(t)$ is the cotrended flux, $c$ is the crowding metric, and $f$ is the flux fraction. The \textit{median} is used here, rather than the \textit{mean}, to make the normalization robust against mathematical outliers. This is important in \textit{Kepler} light curves because transits from planets or eclipsing binaries represent mathematical outliers.


\subsubsection{XI. Goodness Metric}

Characterizing the performance of systematic trend removal in PDC can be highly subjective which emphasizes the difficulty in the cotrending task. A human can peruse a selection of light curves and identify poorly cotrended examples to rate performance. Obvious trends and features can be evaluated but subtle issues with the performance will likely be overlooked.  So, in an effort to minimize the subjective analysis, a numerical \emph{Goodness Metric} has been developed. There are three critical aspects of the cotrending performance to consider in the metric; 1) Removal of systematic trends, 2) no introduced noise and 3) preservation of stellar variability.  A successful cotrend occurs when all three conditions are met, so one cannot rely solely on target-to-target correlation, for example, to determine acceptable performance. Overfitting, and hence flattening of stellar features, results in zero correlation but would not be considered a ``good'' correction. The \emph{Goodness Metric} calculates the cotrending performance in each of the three components using the following methods which were developed experimentally to agree with observations of critical features.

\newcommand{\mrm}[1]{\mathrm{#1}}
Removal of systematic trends is evaluated by the target-to-target correlation. If the correlation is near zero then in principle no systematics can remain\footnote{The root sum-square mean of off-diagonal correlations of white Gaussian noise is exactly  $1/\sqrt{N_{\mrm{sample}}}$ so even with no systematics the off-diagonal correlation matrix will not be exactly zero.}. The Pearson correlation, $C$, is computed over the entire module output and the correlation goodness, $G_{c}$, is computed as the arithmetic mean of the cube of the absolute value of the correlation between the target under study and all other targets,
\begin{equation}
G_{C,i} = \alpha_{C} \frac{1}{N} \sum_{j \ne i}\left({|C_{ij}|}^{3}\right) \quad,
\end{equation}
where $i$ refers to the target under study and $j$ is summed over all other targets. $N$ is the total number of targets. The purpose of the cube is to over-emphasize any strong correlations. $\alpha_{c}$ is a scaling factor.

Introduced noise is determined by examining the change in the Power Spectrum Density of the noise floor for each light curve before and after the correction,
\begin{eqnarray}
\Delta \mrm{PSD(\mathit{freq})} &=& \frac{\mrm{PSD}\left(\mrm{Nf_{after}}\right)} {\mrm{PSD}\left(\mrm{Nf_{before}}\right)}, \\
G_{N,i} &=& \alpha_{N} \int_{\Delta \mrm{PSD} > 1}  \log \left(\Delta \mrm{PSD(\mathit{freq})}\right) \; d\mathit{freq},
\end{eqnarray}
where $\mrm{Nf}$ is the noise floor of the light curve, here defined as the first differences between adjacent flux values. The noise floor is used because removal of trends can result in large changes in the power spectrum density, but signals close to the Nyquist frequency are dominated by noise. The integral is also only taken about regions where the change in power is greater than one, implying an increase in noise.

The final component, preservation of stellar variability, is the most difficult of the three to numerate. An exact determination would require knowledge of the intrinsic signal. Since it is not possible to make this determination, an estimate is made by assessing a mid-frequency band region where light curve signals are dominated by stellar features. High frequency components are removed via a Savitzky-Golay Filter and the low frequency via a 3rd order polynomial removal. What is left is almost always overwhelmingly dominated by stellar signals\footnote{As the residual is typically only very poorly correlated.}. The exception being the Earth-point recovery regions which can be quite strong, so these regions are masked. The standard deviation of the difference between the resultant mid-frequency light curves before and after the correction, $\tilde{y}_{\mrm{before}}$ and $\tilde{y}_{\mrm{after}}$ respectively, is then computed,
\begin{equation}
G_{V,i} = \alpha_{V} \; \mrm{std} \left( \tilde{y}_{i,\mrm{after}} - \tilde{y}_{i,\mrm{before}}\right)^{2} \sqrt{V_{i}} \quad,
\end{equation}
where $V$ is the variability as calculated by PDC-MAP and is used to emphasize targets with greater variability.

The above three components increase in value (in the range of [0,$\infty$)\,) as the performance becomes poorer. But we wish for the ``goodness'' to vary between 0 and 1, 1 being perfect goodness, and so each component is inverted,
\begin{equation}
G'_{k,i} = \frac{1}{G_{k,i} + 1} \quad.
\end{equation}
The total goodness is then the geometric mean of the three components,
\begin{equation}
G_{i} = \sqrt[3]{G'_{C,i} G'_{N,i} G'_{V,i}} \quad.
\end{equation}
The three weighting parameters are used to adjust the relative emphasis of the three components. They have been empirically tuned to agree with observations of cotrending performance, with the weighting parameters being $\alpha_{C} = 12.0$, $\alpha_{N} = 1.0E-4$, and $\alpha_{V} = 1.0E4$.

The above computed \emph{Goodness Metric} has been experimentally developed. It is used when reviewing data products and as an aid to data users to give an estimate of how well to trust the cotrending performance. For the future it is planned to incorporate the \emph{Goodness Metric} as a Lagrange multiplier in the MAP Posterior PDF and to iteratively improve the MAP fit.


\subsection{Outputs of PDC}
The output of PDC is a data structure containing the corrected light curves. If Harmonics Removal was performed (step V), a second set of corrected light curves without harmonic content is exported as well. In addition to these flux time series PDC outputs data processing details such as the target and cadence indices of all SPSDs that were identified, the location of outliers, and diagnostic information such as the Bayesian prior weighting for each target. Of these data, TPS uses the corrected flux time series with filled gaps. The corrected flux times series, cotrending basis vectors, and additional processing data such as the Goodness Metric for each target are exported to the MAST.


\subsection{Processing Times}
\begin{figure}[t]
\centering
\epsscale{0.4}
\plotone{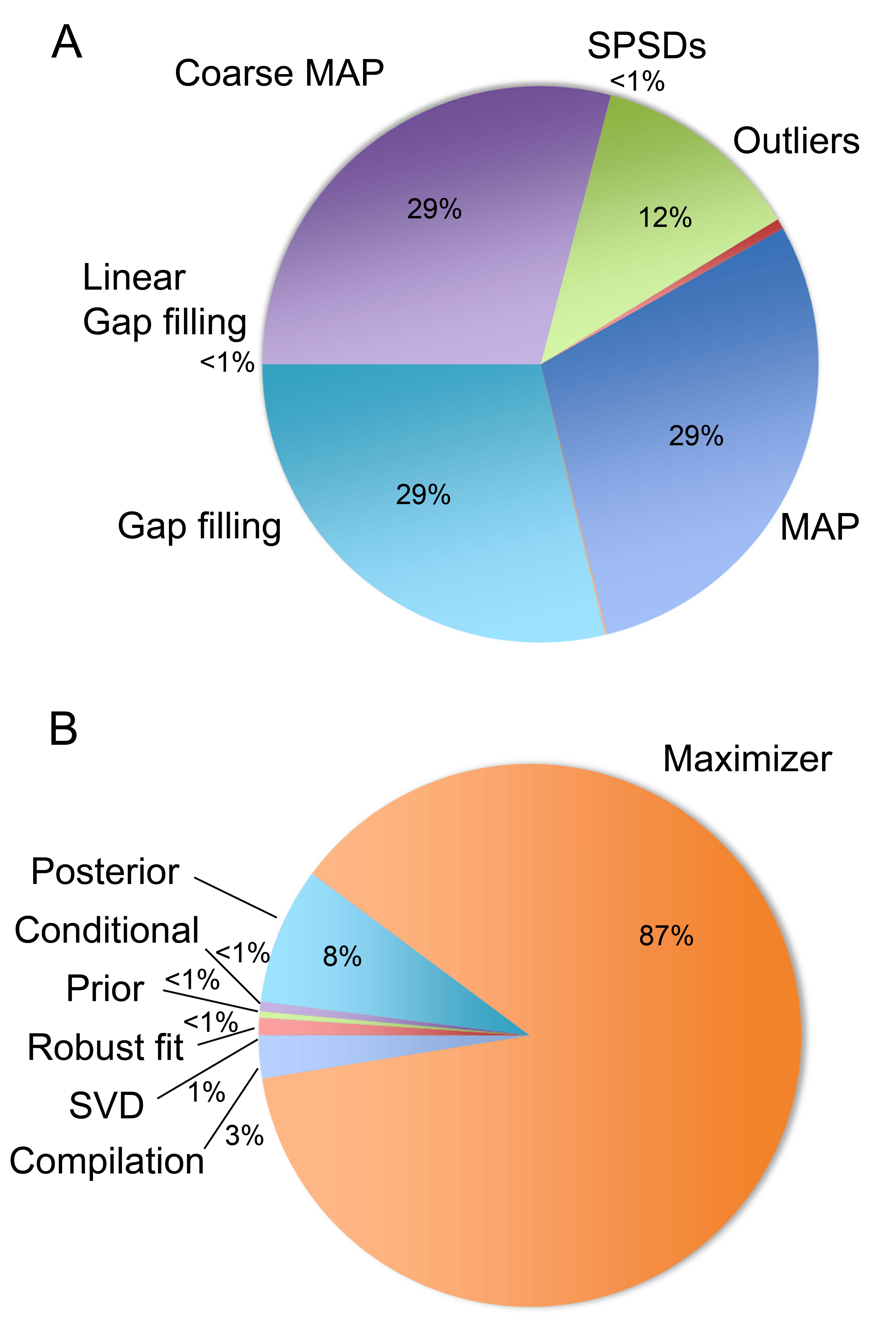}
\caption{\label{fig:processingtimes}Fractions of total processing time for PDC (panel A), and for the MAP part of PDC (panel B).}
\end{figure}

The unit of work for PDC (long-cadence) is one channel for one quarter, so typically 1000--3500 targets with $\approx$4500 data points each. Processing of \textit{Kepler} data is performed on the \textit{Kepler} SOC computer clusters at NASA Ames Research Center and on the Pleiades supercomputer. The relative processing times of the different PDC components is displayed in Figure~\ref{fig:processingtimes}A and shows that the MAP-fit and the final gap filling are the computationally most expensive operations. Within MAP (Fig.~\ref{fig:processingtimes}B), the maximization of the posterior PDF consumes the most time. Processing of a typical module output with 1000--3500 targets with PDC-MAP takes between 20 and 30 hours on a modern 3\,GHz CPU with four cores.

\section{Summary and Conclusions}
We have presented the new (SOC 8.0) Presearch Data Conditioning (PDC) module of the \textit{Kepler} Science Processing pipeline. The task of the PDC module is to correct the different systematic errors in the \textit{Kepler} light curves that otherwise occlude planet transits and other astrophysical signatures. Errors that are corrected by PDC include systematic trends in the data (usually resulting from difference velocity aberration, temperature drift, and pointing errors due to vibrations), flux discontinuities (caused by cosmic rays or solar energetic particles), outliers, and scalar corrections to account for the finite point-spread-function of the targets. Using a Bayesian Maximum A Posteriori approach to remove correlated trends from the light curves, the new PDC (PDC-MAP) has a significantly improved performance as compared with the previous version (PDC-LS) that used a least squares fitting approach with ancillary engineering data~\mycite{twicken10b}. In particular, PDC-MAP is able to remove systematic errors but at the same time preserve the stellar variability.

Despite the dramatic improvements in data correction of this new PDC module, several further improvements are conceivable and will be implemented in the future. Foremost, the generation of basis vectors for the MAP cotrending could potentially be improved by performing band-splitting on the light curves and thus operating MAP on different scales independently. Further, the MAP cotrending algorithm has several free parameters (e.g.\ the number of basis vectors) that are given as input parameters. Ideally, these parameters would be chosen automatically, for instance by optimizing the Goodness Metric. Another potential improvement would be to have time-dependent corrections for the flux-fraction and the crowding-metric, which in the current version are approximated as scalar corrections for each quarter. Finally, due to limitations described above, PDC-MAP is currently only used for \textit{Kepler} long-cadence light curves, whereas the old PDC-LS is still being used for short-cadence light curves -- although short-cadence data, with its high relevance for asteroseismological studies amongst others, could greatly benefit from the superior performance of PDC-MAP as well. Even though some limitations inherent to short-cadence data (such as the sparsity of targets per channel, prohibiting the generation of proper priors in MAP) render it impossible to directly apply PDC-MAP, as is, to short-cadence data, several options exist to partly circumvent these limitations and extend the applicability of PDC-MAP to short-cadence data, such as using interpolated basis vectors from long cadence processing to perform removal of short-cadence systematic trends.

\acknowledgements{Funding for this Discovery Mission is provided by NASA's Science Mission Directorate. We thank the thousands of people whose efforts made \textit{Kepler's} grand voyage of discovery possible. We especially thank the \textit{Kepler} SOC staff who helped design, build, and operate the \textit{Kepler} Science Porcessing Pipeline for putting their hearts into this endeavor.}

{\it Facilities:} \facility{Kepler}


\bibliographystyle{apj}


\end{document}